\documentclass[aps,twocolumn,pra,showpacs,floatfix]{revtex4}
\usepackage{epsfig}
\usepackage{graphicx}
\usepackage{dcolumn}
\usepackage{amssymb,amsmath}
\usepackage{mathrsfs}
\begin{document}

\newcommand{\Z}{Z_{\rm eff}}
\newcommand{\Zr}{\Z^{(\rm res)}}
\newcommand{\eps}{\varepsilon}
\newcommand{\eb}{\varepsilon _b}

\title{Effect of nuclear quadrupole moment on parity nonconservation in atoms}

\author{ V. V. Flambaum, V. A. Dzuba and C. Harabati}

\affiliation{School of Physics, University of New South Wales, Sydney 2052,
Australia}

\begin{abstract}
Nuclei with spin $I \ge 1$ have a weak  quadrupole moment which
leads to tensor contribution to the parity non-conserving interaction between nuclei and electrons. We calculate this
contribution  for Yb$^+$, Fr and Ra$^+$ and found it to be small. In contrast, in many lanthanides (e.g., Nd, Gd, Dy, Ho, Er, Pr, Sm)
and Ra close levels of opposite parity lead to strong enhancement of the effect making it sufficiently large to be measured.
Another possibility is to measure the PNC transitions between the hyperfine components of the ground state of Bi.
Since nuclear weak charge is dominated by neutrons this opens a way of measuring quadrupole moments of neutron distribution in nuclei.
\end{abstract}

\pacs{31.15.A-,11.30.Er}

\maketitle

\section{Introduction}

Studying parity-nonconservation (PNC) in atoms is a way of testing the standard model at low energy as well as
searching for new physics beyond the standard model (see, e.g.,~\cite{K91,GF04}). The most precise measurements
of the PNC in Cs~\cite{WBCMRTW97} supported by accurate atomic calculations~\cite{D89,BJS90,KPT01,DFG02,PBD09,DBFR12} 
show no significant deviation from the standard model. Atomic PNC experiments may also measure  the nuclear
anapole moments \cite{FK80,FKS84,VMMLF95,WBCMRTW97,FM97} and the ratio of PNC amplitudes for different isotopes which 
is not sensitive to the accuracy of atomic calculations ~\cite{DFK86,TDFSYB09}.

Atomic PNC measurements can also be used to study the neutron distribution in nuclei. Several studies looked at the effect
of {\em neutron skin} (the difference  in radius of the proton and neutron distributions) on the PNC in atoms and 
demonstrated that it can give a small but measurable contribution to the PNC
amplitude (see e.g. ~\cite{BDF09}). The study of the neutron distribution should help to establish the equation of state for the 
nuclear matter and properties of neutron stars including the mass boundary for the stability of neutron stars (the neutron 
repulsion at short distances prevents collapse of a neutron star to a black hole).  

In present paper we provide a theory for  a different method to study the neutron distribution in atomic PNC experiments. 
It was noted in Ref. ~\cite{FS78} (see also~\cite{K91}) that the nuclear quadrupole moment induces a tensor PNC weak 
interaction between the nucleus and electrons in atoms and molecules. 
In Ref.~\cite{KP91} it was shown that the combined action of the weak charge and the quadrupole hyperfine interaction 
produces a similar effect but of a significantly smaller amplitude. Note however, that such effect may be enhanced 
if there are close levels mixed by the quadrupole hyperfine interaction.

In Ref.~\cite{F16} it was argued that the tensor  
effects of the weak quadrupole moments are strongly enhanced for deformed nuclei and may get a significant additional 
enhancement due to the close atomic and molecular levels of opposite parity with a  difference of the electron angular 
momenta $|J_1-J_2| \le 2$. These selection rules are similar to that for the effects of the time reversal (T) and parity (P) 
violating nuclear magnetic quadrupole moment (MQM). Therefore, nuclei, molecules and molecular levels  suggested 
for the MQM search in Ref. \cite{FDK14},  for example,  $|\Omega |=1$ doublets in the molecules 
$^{177}$HfF+, $^{229}$ThO, $^{181}$TaN will also have enhanced effects of the weak quadrupole.

Differences in the selection rules  for the  scalar weak charge  ($J_1-J_2=0$), vector anapole moment  ($|J_1-J_2| \le 1$) 
and the tensor weak quadrupole moment ($|J_1-J_2| \le 2$) or the  difference in  the dependence of the PNC effect on 
the hyperfine components of an  atomic transition if more than one operator contribute,  allows one to separate the 
contribution of the weak quadrupole. 

The weak charge of the neutron (-1) exceeds the weak charge of the proton (0.08) by more than an order of magnitude. 
Therefore, the measurements of the PNC effects produced by the weak quadrupole moment  allows one to measure the 
quadrupole moments of the neutron distribution in nuclei. 

In present paper we perform the relativistic many-body calculations of the weak quadrupole effects in atoms of experimental interest.
     
 \section{Theory}

An effective single-electron interaction operator that is responsible for parity nonconservation (PNC) in atom is given by
\begin{align}
\label{eq:tensor_int} 
h_{PNC}=-\frac{G_F}{\sqrt{2}}\gamma_5[ ZC_{1p}\rho_{0p}(r) + N C_{1n}\rho_{0n}(r)]\nonumber\\
-\frac{G_F}{\sqrt{2}}\gamma_5 Y_{20}[ ZC_{1p}\rho_{2p}(r) + N C_{1n}\rho_{2n}(r)].
\end{align}
where $G_F\approx 2.2225 \times 10^{-14}$ in atomic unit (a.u.) is the Fermi constant, the Dirac matrix $\gamma_5$ is defined as in Ref. \citep{K91},  $Z$ and $N$ are the number of protons and neutrons, the coefficients $2C_{1p}=(1-4\sin^2\theta_W) \approx 0.08$, $2C_{1n}=-1$ are the proton and neutron weak charges,  $\rho_p(\mathbf{r})\approx \rho_{0p}(r)+\rho_{2p}(r)Y_{20}(\theta,\phi)$ and $\rho_n(\mathbf{r})\approx \rho_{0n}(r)+\rho_{2n}(r)Y_{20}(\theta,\phi)$ are proton and neutron densities in a nucleus normalized to unity, $\int\rho(r)~d^3r=1$. We have taken into account that if the nuclear spin has the maximal (or any fixed) projection on the $z$ axis the quadrupole part of the density is proportional to  $Y_{20}(\theta,\phi)$. 

Below we will concentrate  on the neutron contribution since the proton contribution to the weak charge is small and may be treated as a correction. Therefore, to simplify the formulae  we assume that the spherical part of the proton  density distribution is equal to that for neutrons:  $\rho_{0p}=\rho_{0n}=\rho_0$. Anyway, the neutron skin is small.

If we assume that $\rho_{2n}(r)=K_{n}\rho_{0}(r)$, the proportionality constant can be expressed in terms of the  quadrupole moment
$Q_n=Q_{zz}=N \int(3z^2-r^2)\rho(\mbox{{\boldmath $r$}})~ d^3r\nonumber$: $K_{n}=\sqrt{5}Q_{n}/(4N\sqrt{\pi}\int\rho_{0} r^4~dr)$ and the tensor part of the  weak interaction is 
\begin{align}
\label{eq:quad_int}
h_Q=-\frac{G_F}{2\sqrt{2}}\gamma_5 Y_{20}\rho_{0}
\frac{\sqrt{5 \pi }Q^{TW}}{\langle r^2 \rangle}\,,
\end{align}
where $Q^{TW} =2C_{1n}Q_n +  2C_{1p}Q_p=-Q_n+0.08 Q_p$ is the weak quadrupole moment,    $\langle r^2\rangle =4 \pi \int\rho_{0} r^4~dr \approx 3R_N^2/5$ is  the mean squared nuclear radius, $R_N$ is the nuclear radius. The quadrupoles $Q_p $ of the proton distribution in nuclei are measured and tabulated in the literature. The neutron quadrupoles $Q_n$ have never been measured. In deformed nuclei  $Q_n \approx (N/Z) Q_p$.
 
In the electromagnetic transitions between the hyperfine components the nuclear spin projection changes, and we should present the  PNC interaction Hamiltonian in terms of the irreducible tensor components:
\begin{align}
\label{eq:quad_int}
h_Q=- \frac{5 G_F }{2\sqrt{2} \langle r^2\rangle}\sum_q(-1)^q T_q^{(2)}Q^{TW}_{-q}\,,
\end{align}
where $T_q^{(2)}=C_q^{(2)}\gamma_5\rho_0(r)$ is the electronic part of the operator,  $Y_{20}=\sqrt{5/(4\pi)}C_0^{(2)}$ and for the second rank tensor $Q^{TW}=2Q^{TW}_0$. 
 
The PNC electric dipole amplitude between states ($|i\rangle\rightarrow |f\rangle$) with the same parity  due to the tensor weak interaction is:
\begin{align}
\label{eq:quad_PNC}
E^{i\rightarrow f}_{PNC}= \sum_n\left [\frac{\langle f|\mbox{{\boldmath $d$}}|n\rangle\langle n|h_Q|i\rangle}{E_i-E_n}+\frac{\langle f|h_Q|n\rangle\langle n|\mbox{{\boldmath $d$}}|i\rangle}{E_f-E_n}\right ]
\end{align}
where $|a\rangle\equiv |J_aF_aM_a\rangle$ is a hyperfine sate and $\mbox{{\boldmath $d$}}=-e\sum_i\mbox{{\boldmath 
$r_i$}}$ is the electric dipole operator, {\boldmath$F=J+I$} is the total angular momentum of an atom, $J$ is the electron 
angular momentum and $I$ is the nuclear spin. More detailed formulae are presented in the appendix.

In performing numerical calculations we follow our earlier work~\cite{DF11} on spin-dependent PNC in single-valence-electron 
atoms. We include nuclear anapole moment contribution as well, so that in most of cases the total PNC amplitudes 
consist of three terms, the spin-independent contribution due to weak nuclear charge, the anapole moment contribution, 
and the weak quadrupole moment contribution. This allows us to fix relative sign of all three terms.
Random phase approximation (RPA) is used for all operators of external fields, including the PNC operators 
and the electric dipole operator. Brueckner orbitals are used to include the core-valence correlations (see \cite{DF11}
for details).

We also use analytical estimations to check numerical results and their uncertainty.  
To do this we use the radial wave functions near the nucleus  from Ref. \cite{K91}:
\begin{align}
\label{eq:f} 
f_{n\kappa}=\frac{\kappa}{|\kappa|}(\kappa-\gamma)\left(\frac{Z}{a_0^3\nu^3}\right )^{1/2}\frac{2}{\Gamma (2\gamma+1)}\left(\frac{a_0}{2Z}\right )^{1-\gamma}r^\gamma
\end{align}
\begin{align}
\label{eq:g} 
g_{n\kappa}=\frac{\kappa}{|\kappa|}Z\left(\frac{Z}{a_0^3\nu^3}\right )^{1/2}\frac{2}{\Gamma (2\gamma+1)}\left(\frac{a_0}{2Z}\right )^{1-\gamma}r^\gamma
\end{align}
where $\gamma=\sqrt{\kappa^2-Z^2\alpha^2}$, $a_0$ is Bohr radius,  $\nu_n^2=-1/(2\epsilon_n)$ is 
the effective principle quantum number, $\epsilon_n$ is the orbital energy in a.u. and  $\Gamma(x)$ is the Gamma function.

Analytical and numerical results agree on the level of 30\% or better. The accuracy of the numerical results is few per cent 
for Fr and Ra$^+$ and $\sim $ 30\% for Yb$^+$. A detailed analysis of accuracy of the calculations can be found in Ref.~\cite{DF11}.
We belive that the accuracy of our present calculations is the same as in Ref.~\cite{DF11}.

 \section{Results and Discussion}
\subsection{The $s-s$ and $s-d$ transitions}
\label{s:sd}


\begin{table}
\caption{PNC amplitudes $\langle a,F_1|E^{\rm PNC}_z|b,F_2 \rangle$ ($z$-components) 
for the $s-s$ and $s-d$ transitions in $^{173}$Yb$^+$ ($I=7/2, \ Q_W= -96.84$),
$^{223}$Fr ($I=3/2, \ Q_W= -128.25$), and $^{223}$Ra$^+$ ($I=3/2, \ Q_W= -127.2$).
Weak nuclear charge ($Q_W$), nuclear anapole moment ($\varkappa$), and neutron quadrupole
moment ($Q_n$) contributions are presented. The unit for $Q_n$ is barn (1b = $10^{-24}$cm$^2$).}
\label{t:sdpnc}
\begin{ruledtabular}
\begin{tabular}{l ccr}
\multicolumn{1}{c}{Isotope/} & 
$F_1$ & $F_2$ & \multicolumn{1}{c}{PNC amplitude} \\
\multicolumn{1}{c}{Transition} &&& \multicolumn{1}{c}{$10^{-10} iea_0$} \\
\hline
$^{173}$Yb$^+$  &1 & 2 &  $-0.41\times [1-0.022\varkappa - 7.5\times 10^{-6}Q_n]$ \\
$5d_{3/2} - 6s$   & 2 & 2 &  $-0.53\times[1-0.016\varkappa- 2.7\times 10^{-6}Q_n]$ \\
& 2 & 3 &  $-0.17\times[1-0.005\varkappa+ 1.4\times 10^{-5}Q_n]$ \\
& 3 & 2 &  $ 0.28\times[1+0.007\varkappa-1.1\times 10^{-5}Q_n]$ \\
& 3 & 3 &  $-0.48\times[1+0.004\varkappa+2.1\times 10^{-6}Q_n]$ \\
& 4 & 3 &  $ 0.37\times[1-0.016\varkappa+ 2.7\times 10^{-6}Q_n]$ \\

$^{173}$Yb$^+$  & 1 & 2 &  $-6.8\times 10^{-4}\varkappa + 4.8\times 10^{-6}Q_n$ \\
$5d_{5/2} - 6s$    & 2 & 2 &  $-1.4\times 10^{-3}\varkappa+ 1.1\times 10^{-5}Q_n$ \\
& 2 & 3 &  $-4.6\times 10^{-4}\varkappa- 3.1\times 10^{-6}Q_n$ \\
& 3 & 2 &  $ 9.4\times 10^{-4}\varkappa-8.8\times 10^{-6}Q_n$ \\
& 3 & 3 &  $-1.6\times 10^{-3}\varkappa-8.5\times 10^{-6}Q_n$ \\
& 4 & 3 &  $1.2\times 10^{-3}\varkappa+ 4.1\times 10^{-6}Q_n$ \\

$^{223}$Fr & 1 & 1 &  $-0.31\times [1-0.023\varkappa + 3.4\times 10^{-5}Q_n]$ \\
$7s - 8s$    & 1 & 2 &  $0.54\times[1+0.17\varkappa+ 6.7\times 10^{-6}Q_n]$ \\
& 2 & 1 &  $0.54\times[1-0.16\varkappa+ 6.7\times 10^{-6}Q_n]$ \\
& 2 & 2 &  $ 0.63\times[1-0.014\varkappa-6.7\times 10^{-6}Q_n]$ \\

$^{223}$Fr & 1 & 0 &  $-4.5\times [1-0.026\varkappa - 1.1\times 10^{-5}Q_n]$ \\
$7s - 6d_{3/2}$ & 1 & 1 &  $-5.0\times[1-0.026\varkappa- 4.3\times 10^{-6}Q_n]$ \\
& 1 & 2 &  $ 3.9\times[1+0.026\varkappa- 8.9\times 10^{-6}Q_n]$ \\
& 2 & 1 &  $-1.7\times[1+0.016\varkappa+7.8\times 10^{-6}Q_n]$ \\
& 2 & 2 &  $-4.5\times[1+0.016\varkappa+4.8\times 10^{-6}Q_n]$ \\
& 2 & 3 &  $ 4.5\times[1-0.015\varkappa+ 2.2\times 10^{-6}Q_n]$ \\

$^{223}$Fr        &1 & 1 &  $ 4.9\times 10^{-3}\varkappa + 1.0\times 10^{-4}Q_n$ \\
$7s - 6d_{5/2}$ & 1 & 2 &  $-5.8\times 10^{-3}\varkappa- 1.2\times 10^{-4}Q_n$ \\
                          & 2 & 1 &  $ 1.7\times 10^{-3}\varkappa- 9.1\times 10^{-6}Q_n$ \\
                          & 2 & 2 &  $ 6.7\times 10^{-3}\varkappa-4.0\times 10^{-5}Q_n$ \\
                          & 2 & 3 &  $-7.2\times 10^{-3}\varkappa+5.0\times 10^{-5}Q_n$ \\

$^{223}$Ra$^+$ & 1 & 0 &  $-3.0\times [1-0.025\varkappa - 9.5\times 10^{-6}Q_n]$ \\
$7s - 6d_{3/2}$  & 1 & 1 &  $-3.4\times[1-0.022\varkappa- 5.2\times 10^{-6}Q_n]$ \\
                             & 1 & 2 &  $ 2.6\times[1+0.016\varkappa- 8.9\times 10^{-6}Q_n]$ \\
                             & 2 & 1 &  $-1.2\times[1+0.0003\varkappa+1.4\times 10^{-5}Q_n]$ \\
                             & 2 & 2 &  $-3.0\times[1+0.0062\varkappa+2.4\times 10^{-6}Q_n]$ \\
                             & 2 & 3 &  $ 3.0\times[1-0.015\varkappa+ 1.9\times 10^{-6}Q_n]$ \\

$^{223}$Ra$^+$ &1 & 1 &  $ 1.4\times 10^{-3}\varkappa + 6.2\times 10^{-5}Q_n$ \\
$7s - 6d_{5/2}$  & 1 & 2 &  $-1.6\times 10^{-3}\varkappa- 7.8\times 10^{-5}Q_n$ \\
                          & 2 & 1 &  $ 4.7\times 10^{-4}\varkappa- 1.0\times 10^{-5}Q_n$ \\
                          & 2 & 2 &  $ 1.8\times 10^{-3}\varkappa-3.5\times 10^{-5}Q_n$ \\
                          & 2 & 3 &  $-1.9\times 10^{-3}\varkappa+2.9\times 10^{-5}Q_n$ \\
\end{tabular}
\end{ruledtabular}
\end{table}

Calculated PNC amplitudes between different hfs components of $s$ and $d$ states of 
$^{173}$Yb$^+$, $^{223}$Fr, and $^{223}$Ra$^+$ are presented in Table~\ref{t:sdpnc}.
The amplitudes consist of three contributions, the spin-independent contribution
due to nuclear weak charge $Q_W$, the contribution of the nuclear anapole moment $\varkappa$,
and the contribution of the neutron quadrupole moment $q$. We have chosen these atoms because 
they are considered for the PNC measurements (see, e.g.~\cite{YbPNC,FrPNC1,FrPNC2,RaPNC}) and because
some isotopes of these atoms have deformed nucleus and therefore large quadrupole moments for
both proton and neutron distributions. Electric quadropole moments ($Q_p$) are known and 
tabulated~\cite{Qp}. The values for considered isotopes are $Q_p(^{173}{\rm Yb}) = 2.80(4) b$,
$Q_p(^{223}{\rm Fr}) = 1.17(2) b$, and $Q_p(^{223}{\rm Ra}) = 1.25(7) b$.
Using estimate $Q_n \approx (N/Z) Q_p$ we see that the largest contributions of the neutron
quadrupole term to the PNC amplitude is $\sim 10^{-4}$ of the spin-independent contribution.
This is relatively small value which probably means that one should look for enhancement 
factors, such as, e.g. close states of opposite parity. The atoms considered above do not have such
enhancement. They were originally chosen for the measurements of the spin-independent PNC. 
They have large $Z$ (PNC scales as $\sim Z^3$) and relatively simple electron structure
(one electron above closed shells) which allow for accurate interpretation of the measurements. 
The study of neutron quadrupole moments needs different criteria for choosing the objects for
measurements. One could search, e.g. for close states of opposite parity with $\Delta J=2$.
Such states can be only mixed by the neutron quadrupole moment and PNC amplitudes 
involving such states can be enhanced to the measurable level by small energy intervals.
Note also that high accuracy of the calculations is not needed at this stage. Therefore,
promising candidates can probably be found in atoms with dense spectra such as atoms
with open $d$ or $f$ shells. Molecules can be good candidates too. 

\subsection{hyperfine transitions}
\label{s:hfs}
\begin{table}
\caption{Nuclear anapole and neutron quadrupole contributions to the PNC transition
between hfs components of the ground state of $^{209}$Bi ($I=9/2 \ Q_W = -118.65$).}
\label{t:Bihfs}
\begin{ruledtabular}
\begin{tabular}{ccr}
$F_1$ & $F_2$ & \multicolumn{1}{c}{PNC amplitude} \\
&& \multicolumn{1}{c}{$10^{-10} iea_0$} \\
\hline
 3 & 4 &  $ -2.0\times 10^{-4}\varkappa+ 2.6\times 10^{-6}Q_n$ \\
 4 & 5 &  $ -2.4\times 10^{-4}\varkappa+7.6\times 10^{-7}Q_n$ \\
 5 & 6 &  $-2.1\times 10^{-4}\varkappa-1.8\times 10^{-6}Q_n$ \\

\end{tabular}
\end{ruledtabular}
\end{table}

Similar to the anapole moment contribution, the neutron quadrupole moment can lead to PNC
transition between different hyperfine components of the same state. However, there are further
restrictions due to higher rank of the operator. Since quadrupole moment is the rank 2 operator, the minimum value 
of the total angular momentum $J$ of the atomic state to have non-zero contribution is $J=1$. For 
a single-valence-electron atom the minimum value is $J=3/2$. This means that the effect is zero
in the ground state of all atoms considered above. Therefore, we consider Bi atom  instead
for which first measurements of atomic PNC were performed~\cite{BiPNC}.
The results of the calculations are presented in Table~\ref{t:Bihfs}. Using estimations $Q_n \approx (N/Z)Q_p \approx -0.9b$,
and $\varkappa({\rm Bi}) \sim 0.1$~\cite{Bikappa} we see that the neutron quadrupole 
contribution is only about one order of magnitude smaller than the anapole contribution.

\subsection{Close levels of opposite parity in lanthanoids}

\begin{table*}
\caption{Ground states and pairs of close states of opposite parity with $\Delta J=2$ in some lanthanide atoms
and corresponding PNC amplitudes.}
\label{t:dj2}
\begin{ruledtabular}
\begin{tabular}{lll ddcc cl}
\multicolumn{1}{c}{Atom/} & 
\multicolumn{2}{c}{States} &
\multicolumn{1}{c}{$J$} & 
\multicolumn{1}{c}{$E$} & 
\multicolumn{1}{c}{$\tau$} & 
\multicolumn{1}{c}{$\Delta E$} &
\multicolumn{1}{c}{$h_Q$} & 
\multicolumn{1}{c}{$|E_{\rm PNC}|$} \\

\multicolumn{1}{c}{$Q_p(b)$} & & & & 
\multicolumn{1}{c}{ (cm$^{-1}$)} & & 
\multicolumn{1}{c}{ (cm$^{-1}$)} &
\multicolumn{1}{c}{ m.e. } & 
\multicolumn{1}{c}{ (a.u.) }  \\
\hline

$^{143}$Nd   & 4f4.6s2                                  &  5I      &   4.0 & 0.0 & & & \\
-0.61(2)         & 4f3.(4I*).5d2.(3F) (6L*).6s     &  7L*   &   5.0 &   11108.813 &   1 $\mu$s & &$\langle 4f6s|h_Q|5d^2 \rangle$ & $1.1 \times 10^{-13}$  \\
                     & 4f4.6s2                                  &  5F     &   3.0 &   11118.466 & 338 $\mu$s &   9.652 & & \\
        & 4f4.(5I).5d.6s.(3D)           &  7G     &   2.0 &   11990.020 & 273 $\mu$s & &$\langle 4f|h_Q|6s \rangle$ & $5.4 \times 10^{-13}$  \\
                     & 4f3.(4I*).5d.6s2                &  *         &   4.0 &   11992.388 &   1 $\mu$s &   2.368 & &\\
&&&&&&&& \\
$^{155}$Gd & 4f7.(8S*).5d.6s2                &  9D*    &   2.0 & 0.0 & & &\\
1.27(3) & 4f7.(8S*).5d2.(3P) (10P*).6s    &  9P*    &   3.0 &   15173.639 & 407 ns     &  & $\langle 5d|h_Q|6p \rangle$ & $3.9 \times 10^{-11}$ \\
   & 4f7.(8S*).5d (9D*).6s.6p.(3P*)  &  11F             &   5.0 &   15174.000 &   2 $\mu$s &   0.361 & &\\
   & 4f7.(8S*).5d2.(1D) (8D*).6s     &  9D*              &   6.0 &   17906.736 & 655 ns     &  & $\langle 5d|h_Q|6p \rangle$ & $4.3 \times 10^{-12}$ \\
   & 4f7.(8S*).5d (9D*).6s.6p.(3P*)  &  11F             &   8.0 &   17909.943 &   7 $\mu$s &   3.207 & &\\
&&&&&&&& \\
$^{161}$Dy  & 4f10.6s2                    &  5I         &   8.0 & 0.0 & & & \\
2.51(2) & 4f10.(5I$<8>$).5d.6s.(3D)    & 3[9]       &  10.0 &  18462.650 & 11 $\mu$s &  & $\langle 4f|h_Q|5d \rangle$ & $2.0 \times 10^{-12}$ \\
   & 4f9.(6H*).5d2.(3F) (8G*).6s       & 9G*       &   8.0 &   18472.711 & 819 ns     &  10.061 & & \\
&&&&&&&& \\
$^{165}$Ho & 4f11.6s2                        &             4I*    &   7.5 & 0.0 & & & \\
3.58(2) & 4f11.(4I*).5d.6s.(3D)           &             *      &   6.5 &   20493.770 & 771 ns     &  & $\langle 4f|h_Q|5d \rangle$ & $5.8 \times 10^{-12}$  \\
   & 4f10.(5I).5d2.(3F) (7H).6s      &             8H     &   8.5 &   20498.730 &   1 $\mu$s &   4.961 & & \\
   & 4f11.(4I*$\langle 13/2\rangle$).6s.6p.(3P*$\langle 1\rangle$) &         (13/2,1)   &   6.5 &   22157.859 & 512 ns     &  & $\langle 6p|h_Q|5d \rangle$ & $1.9 \times 10^{-9}$  \\
   & 4f11.(4I*).5d.6s.(3D)           &             *      &   4.5 &   22157.881 &   3 $\mu$s &   0.021 & & \\
&&&&&&&& \\
$^{167}$Er & 4f12.6s2                        &         3H         &   6.0 & 0.0 & & & \\
3.57(3)  & 4f11.(4I*).5d.6s.6p             &                    &   7.0 &   25861.232 & 567 ns     &  & $\langle 6p|h_Q|5d \rangle$ & $1.8 \times 10^{-11}$  \\
   & 4f11.(4I*).5d2.(3P) (6I*).6s    &         7I*        &   9.0 &   25863.453 & 507 ns     &   2.221 & & \\
\end{tabular}
\end{ruledtabular}
\end{table*}
As we discussed above the quadrupole PNC contributions is at least four orders of magnitude smaller than the scalar one.
This makes it hard to measure and one should look for enhancement factors. Strong enhancement can take place when
a pair of states of opposite parity is separated by a small energy interval. Such pairs can be found in lanthanoid atoms.
For atoms considered above typical energy denominator (see formula \ref{eq:quad_PNC}) is $\sim$ 10,000~cm$^{-1}$.
Therefore, for the quadrupole contribution being similar in value with the scalar one we need to look for energy 
intervals between states of opposite parity $\sim$ 1~cm$^{-1}$.  We shall consider close states with the difference in the
value of the total angular momentum $\Delta J=1,2$. The opposite parity states with $\Delta J=2$ can only be mixed 
by weak quadrupole making it the only contribution to the PNC amplitude. This is a clear case for the weak
quadrupole study. In contrast, the states with $\Delta J=1$ can be mixed by both, weak quadrupole and nuclear anapole.
We consider these two cases separately.

\begin{figure*}
\epsfig{figure=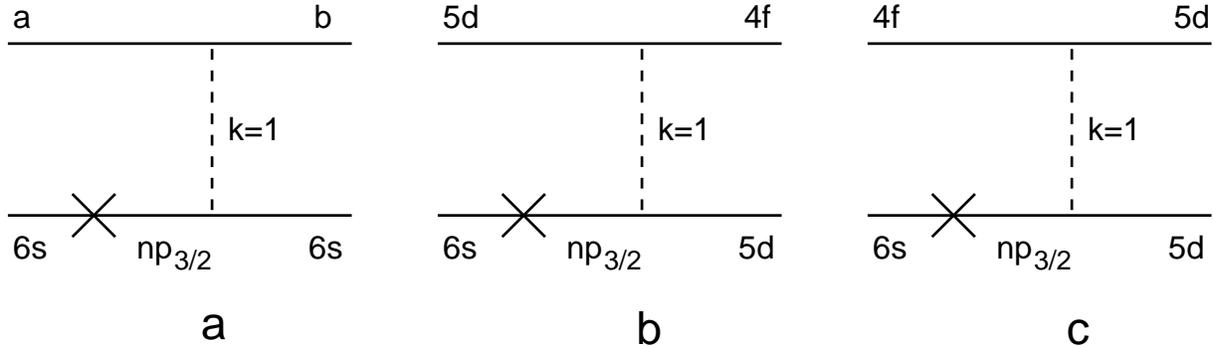,scale=0.9}
\caption{Coulomb corrections to the weak matrix elements. Cross stands for the weak quadrupole interaction; summation over
complete set of $np_{3/2}$ states is assumed. a. The $6p-5d$ or $5d-4f$ matrix elements; b. The  $6s-4f$ matrix element; 
c. The two-electron matrix element between the $4f6s$ and $5d^2$ states (e.g., in Nb).}
\label{f:F1}
\end{figure*}

\begin{table}
\caption{Matrix elements of the neutron quadrupole operator $h_Q$ and Coulomb corrections to them (a.u.).
RHF stands for relativistic Hartree-Fock, RPA is the random-phase approximation. Numbers in square brackets 
stand for powers of ten.}
\label{t:qme}
\begin{ruledtabular}
\begin{tabular}{lrrr}
\multicolumn{1}{c}{Transition} & 
\multicolumn{2}{c}{$\langle a |H^Q| b\rangle$} &
\multicolumn{1}{c}{$\langle \tilde{6s},a |\frac{e^2}{|r_1-r_2|}| 6s,b\rangle$} \\
& \multicolumn{1}{c}{RHF} &\multicolumn{1}{c}{RPA} & \\ 
\hline
$4f_{5/2} - 6s_{1/2}$ & 2.24[-19]& 2.50[-19] & 7.78[-17] \\
$4f_{5/2} - 5d_{3/2}$& 5.83[-25] & 2.95[-19]&  2.14[-16] \\
$4f_{5/2} - 5d_{5/2}$& -1.89[-25] & 2.54[-18]&  -5.50[-17] \\

$4f_{7/2} - 5d_{5/2}$& 5.57[-30] & 4.60[-18]&  2.54[-16] \\

$6p_{3/2} - 6s_{1/2}$& 2.11[-16] & 5.62[-16] & 4.40[-16] \\

$6p_{1/2} - 5d_{3/2}$& 5.00[-18] & 1.46[-17] & -3.46[-16] \\

$6p_{3/2} - 5d_{3/2}$& 4.30[-19] & -1.72[-17] & 1.39[-16] \\
$6p_{3/2} - 5d_{5/2}$& 4.27[-23] & -4.15[-17] & 4.27[-16] \\
  
\end{tabular}
\end{ruledtabular}
\end{table}

\subsubsection{Close states with $\Delta J=2$.}

Table \ref{t:dj2} shows some examples of the pairs 
of states for lanthanoid atoms separated by energy interval $\Delta E \le 10$~cm$^{-1}$ and having the values of the
total angular momentum $J$ which differ by 2. The data has been obtained by analysing the NIST databases~\cite{NIST}.
We include only states which seem to be promising for the study of the PNC caused by neutron quadrupole moment.
We excluded atoms where all stable isotopes have small nuclear spin ($I<1$) and thus no quadrupole moment.
We excluded highly excited states and pairs of close states if an electron configuration for at least one state is not known.

Neither scalar nor anapole PNC interactions can mix the states with $\Delta J=2$.
The weak quadrupole is the only contribution to the PNC involving the states. This makes them good candidates for the
study of the neutron quadrupole moments.  If one of the states is connected to the ground state by an electric dipole transition (E1)
one can study the interference between Stark-induced and PNC-induced amplitudes of the transition to the ground state
similar to what was measured in Cs~\cite{WBCMRTW97}. Otherwise, one can study the interference between the hyperfine 
or Stark-induced and the PNC-induced amplitudes of the transition between these two states similar to what was done
for Dy~\cite{DyPNC}. In latter case one needs metastable states. Therefore, we performed estimations of the lifetimes 
of each state in the Table. The estimations are approximate. We consider only E1 transitions, using experimental energies
and assuming that all E1 amplitudes are equal to 1~a.u. The results are presented in Table~\ref{t:dj2}. 

The $E^{\rm PNC}$ amplitude is estimated using the formula
\begin{equation}
E^{\rm PNC}_{ag} \sim c_0 \frac{\langle a|h_Q|b\rangle\langle b|D|g\rangle}{\Delta E} Q_n.
\label{eq:EPNC}
\end{equation}
Here $a$ and $b$ is a pair of the close-energy states, state $g$ is the ground state, $D$ is an operator 
of the electric dipole transition (E1), $c_0$ is angular coefficient (see formula (\ref{eq:Q_PNC}), $Q_n$ is the neutron
quadrupole moment.
For the estimations we assume $\langle b|D|c\rangle=1$~a.u., $c_0=0.1$, $Q_n =(N/Z) Q_p$. The values of the electric
quadrupole moment $Q_p$ are taken from Ref.~\cite{Qp}. 

Estimations of the $\langle a|h_Q|b\rangle$ matrix
elements are more complicated.  Calculations show that all of them apart from only the $s-p_{3/2}$
matrix elements are very sensitive to many-body effects. This is a well-known feature of any short-range interaction
of atomic electrons with the nucleus. The wave functions of states with angular momentum $l>1$ are negligibly small
on the nucleus and $s$ states of other electrons must come into play via many-body effects to make a dominant contribution.
Table~\ref{t:dj2} indicates that we need to deal with the $s-f$, $p-d$, and $d-f$ weak matrix elements which are sensitive 
to many-body effects. Table~\ref{t:qme} shows the values of the weak matrix elements calculated in the relativistic 
Hartree-Fock (RHF) and RPA approximations (we use Gd atom as an example). Taking into account the core polarization via 
the RPA calculations increases the value of most matrix elements by many orders of magnitude. Further increase can be
found if the configuration mixing is taken into account. Configuration mixing brings into play configurations which make 
possible the $6s-np_{3/2}$ contribution to the weak matrix element. Sample diagrams are presented on Fig.~\ref{f:F1}.
Note that the configuration mixing is due to the Coulomb interaction. Therefore, we call corresponding corrections to the weak
matrix elements the Coulomb corrections.
For example, the   Coulomb corrections to the $\langle 5d|h_Q|6p \rangle$ and $\langle 5d|h_Q|4f \rangle$ matrix
elements in Gd, Ho and Er are given by the diagram~Fig.~\ref{f:F1}.a and the correction to the $\langle 6s|h_Q|4f \rangle$ 
matrix element is given by the diagram Fig.~\ref{f:F1}.b. Note that the weak matrix element between first pair of states in Nb
is zero in the single-electron approximation since the states differ by two electron orbitals. In this case the diagram Fig.~\ref{f:F1}.c
is the lowest-order contribution.

We estimate the diagrams (Fig.~\ref{f:F1}) by calculating Coulomb integrals in which one $6s$ wave function is replaced 
by a correction induced by the $h_q$ operator. The correction is calculated in the RPA approximation
\begin{equation}
\label{eq:6sRPA}
	(H^{RHF} - \epsilon_{6s}) \delta \psi_{6s} = - (h_Q + \delta V^{RHF})\psi_{6s}.
\end{equation}
Corresponding Coulomb integrals are 
\begin{eqnarray}
&\langle \tilde{6s},a |r_</r_>^2| 6s,b\rangle & {\rm Fig}.~\ref{f:F1}.a, \nonumber \\ 
 &\langle \tilde{6s},5d |r_</r_>^2| 5d,4f\rangle & {\rm Fig}.~\ref{f:F1}.b, \nonumber \\ 
 & \langle \tilde{6s},4f |r_</r_>^2| 5d,5d\rangle & {\rm Fig}.~\ref{f:F1}.c. \nonumber 
 \end{eqnarray}
Here $r_<=\min(r_1,r_2)$, $r_>=\max(r_1,r_2)$, and $| \tilde{6s}\rangle \equiv \delta \psi_{6s}$.
Calculated values of the Coulomb integrals are presented in the last column of Table~\ref{t:qme}.
Substituting these numbers into (\ref{eq:EPNC}) we get estimations for the PNC amplitudes. 
The results are presented in Table~\ref{t:dj2}. Note that in contrast 
to the amplitudes considered in sections \ref{s:sd} and \ref{s:hfs} the amplitudes here are relatively large. In most of cases they are 
lager than the PNC amplitude in Cs~\cite{WBCMRTW97}. In the case of second pair of close states in Ho, the amplitude is as large as in Yb,
the largest PNC atomic amplitude which has been measured so far~\cite{TDFSYB09}.

In all cases considered above one can measure the transition rate between the two close states of opposite parity 
and study the interference between the PNC amplitude (\ref{eq:EPNC}) and the electric dipole transition induced 
by the hyperfine interaction.  In addition, when one of the states is connected to the ground state by the magnetic dipole (M1)
transition (first pairs of states in Nd, Gd, Ho and Er) or an electric quadrupole (E2) transition (second pair of states in Nd) one can
study the interference between these M1 or E2 amplitudes and the PNC amplitude (\ref{eq:EPNC}) to the ground state.

\subsubsection{Close states with $\Delta J=1$.}

\begin{table*}
\caption{Ground states and pairs of close states of opposite parity with $\Delta J=1$ in some lanthanide atoms and Ra.}
\label{t:Lan}
\begin{ruledtabular}
\begin{tabular}{lll ddcc}
\multicolumn{1}{c}{Atom} & 
\multicolumn{2}{c}{States} &
\multicolumn{1}{c}{$J$} & 
\multicolumn{1}{c}{$E$ (cm$^{-1}$)} & 
\multicolumn{1}{c}{$\tau$} & 
\multicolumn{1}{c}{$\Delta E$ (cm$^{-1}$)} \\
\hline

Pr & $4f3.6s2$                            &          4I*    &   4.5 & 0.0 & \\
Pr & $4f2.(3H).5d.6s2$                    &          4H     &   5.5 &    9675.010 &   6 $\mu$s &  \\
   & $4f3.(4I*).5d.6s.(3D) $              &          6K*    &   6.5 &    9684.240 &   5 $\mu$s &   9.230 \\
Pr &                                    &                 &   6.5 &   10423.680 & 306 $\mu$s &  \\
   & $4f3.(4I*).5d.6s.(3D) $              &          4K*    &   5.5 &   10431.750 &   3 $\mu$s &   8.070 \\
Pr & $4f2.(3H).6s2.6p$                   &          4I*    &   4.5 &   19339.859 & 113 ns     &  \\
   &                                    &                 &   5.5 &   19343.250 & 356 ns     &   3.391 \\
Nd & $4f4.6s2$                            &       5I        &   4.0 & 0.0 & \\
Nd & $4f3.(4I*).5d2.(3F) (6L*).6s$        &       7L*       &   5.0 &   11108.813 &   1 $\mu$s &  \\
   & $4f4.(5I).5d.6s.(3D)$                &       7K        &   6.0 &   11109.167 &  29 $\mu$s &   0.354 \\
Nd & $4f4.(5I).5d.6s.(3D)$                &       7I        &   6.0 &   12917.422 &   7 $\mu$s &  \\
   & $4f3.(4I*).5d.6s2$                   &       5I*       &   7.0 &   12927.232 &   2 $\mu$s &   9.811 \\
Sm & $4f6.6s2$                            &      7F         &   0.0 & 0.0 & \\
Sm & $4f6.(7F).6s.6p.(3P*)$               &      9G*        &   5.0 &   16344.770 & 753 ns     &  \\
   & $4f6.(7F).5d (8D).6s$                &      7D         &   4.0 &   16354.600 &   1 ms     &   9.830 \\
Gd & $4f7.(8S*).5d.6s2$                   &       9D*       &   2.0 & 0.0 & \\
Gd & $4f7.(8S*).5d (9D*).6s.6p.(3P*)$     &       7D        &   3.0 &   19399.840 & 252 ns     &  \\
   & $4f7.(8S*).5d2.(1G) (8G*).6s$        &       9G*       &   2.0 &   19403.104 & 164 ns     &   3.264 \\
Gd & $4f7.(8S*).5d2.(3F) (6F*).6s$        &       5F*       &   3.0 &   20299.869 & 121 ns     &  \\
   &                                    &                 &   2.0 &   20303.801 & 169 ns     &   3.932 \\
Tb & $4f9.6s2$                            &       6H*       &   7.5 & 0.0 & \\
Tb & $4f8.(7F<6>).6s2.6p<1/2> $           &       (6,1/2)*  &   5.5 &   13616.270 & 430 ns     &  \\
   &                                    &                 &   6.5 &   13622.690 &   1 $\mu$s &   6.421 \\
Dy & $4f10.6s2$                           &        5I       &   8.0 & 0.0 & \\
Dy & $4f10.(5I<8>).6s.6p.(3P*<2>)$        &       (8,2)*    &  10.0 &   17513.330 & metastable  &  \\
   & $4f10.(5I<8>).5d.6s.(3D)$            &        3[8]     &   9.0 &   17514.500 &   5 $\mu$s &   1.170 \\
Dy & $4f9.(6H*).5d2.(3P) (8I*).6s$        &          *      &   9.0 &   23271.740 & 697 ns     &  \\
   & $4f10.(5I<7>).5d.6s.(3D)$            &                 &   8.0 &   23280.461 & 556 ns     &   8.721 \\
Dy & $4f10.(5I<7>).5d.6s.(3D)$           &                 &   6.0 &   23333.920 & 526 ns     &  \\
   & $4f9.(6H*).5d2.(3F) (8G*).6s$        &          *      &   7.0 &   23340.119 & 263 ns     &   6.199 \\
Dy & $4f9.(6H*).5d2.(3F) (8F*).6s$        &          *      &   6.0 &   23359.820 & 359 ns     &  \\
   & $4f10.(5I<7>).5d.6s.(3D)$            &                 &   7.0 &   23360.660 & 435 ns     &   0.840 \\
Ho & $4f11.6s2$                           &          4I*    &   7.5 & 0.0 & \\
Ho & $4f10.(5I<6>).5d<3/2>.6s2$           &       (6,3/2)   &   6.5 &   18564.900 & 935 ns     &  \\
   & $4f10.(5I<8>).6s2.6p<1/2>$           &       (8,1/2)   &   7.5 &   18572.279 &   1 $\mu$s &   7.3 \\

Ra & $7s2$             &       1S    &   0.0 & 0.0 & & \\
Ra & $7s6d$           &       3D    &  2.0  &  13993.94  &  metastable    &  \\
      & $7s7p$           &       3P*  &  1.0  &  13999.3569  &  500 ns    &  5.42 \\

\end{tabular}
\end{ruledtabular}
\end{table*}

Close states of opposite parity with $\Delta J=1$ are also important. Here both, the anapole moment and the weak quadrupole 
moments contribute to the PNC effect. Measuring both these moments are equally important. The anapole moment has been
measured for Cs only~\cite{WBCMRTW97}. The limit on the anapole moment of Tl obtained in the PNC measurements~\cite{VMMLF95}
has also been obtained. Measuring more anapole moments may help to extract constants of the weak
interaction between nucleons and to get better understanding of nuclear structure. Measuring PNC effect which has both,
anapole and quadrupole contributions may have some advantages. The effect is expected to be larger while different
dependence of two contributions on the quantum numbers (e.g., on total angular momentum $F$, $\mathbf{F}=\mathbf{J}+\mathbf{I}$)
allows one to separate the contributions.

Table~\ref{t:Lan} shows pairs of opposite parity states of lanthanoids separated by the energy interval $\Delta E <10~{\rm cm}^{-1}$
with values of the total angular momentum $J$ which differ by one. The pairs have been found by analysing the NIST
database~\cite{NIST}. We also included Ra which was studied in Ref.~\cite{Ra-Ginges}.

It is clear that many of the systems listed in Table~\ref{t:Lan} are as good as those considered in the previous section. 
Estimations can be also done in a similar way. The most important parameters defining the value of the PNC amplitude 
are the energy interval between states of opposite parity an the type of the weak matrix element. The values for different
types of weak matrix elements are presented in Table~\ref{t:qme}. The energy intervals are presented in Table~\ref{t:Lan}.
More detailed study of the PNC amplitudes for all systems listed in Table~\ref{t:Lan}  goes beyond the scope of 
present work.  The analysis can be done for a particular system which is of the greatest interest to experimentalists. 
In our veiw there are many systems which look very promising but require careful consideration from the experimental point of view.

\section{Conclusion} 
 
 We argue that the measuring PNC in atoms can be used to study the neutron distribution in nuclei via measuring the parity-nonconserving
 weak quadrupole moment. The effect is small in atoms which have been already used to study PNC. However, a strong enhancement 
 due to close states of opposite parity can be found in lanthanoids and in Ra. Here the neutron quadrupole moments can be studied together
 with the nuclear anapole moments. There many systems where the weak quadrupole moment is the only enhanced 
 contribution to the PNC effect.
 The enhancement is sufficiently strong to make the prospects of the measurements to be very realistic.

\acknowledgments

This work was funded in part by the Australian Research Council.  

\appendix
\section{Matrix elements}
The projection $M$ dependence of the amplitude can be factorized by using the Wigner-Eckart theorem:
\begin{align}
\label{eq:quad_reduced}
E^{i\rightarrow f}_{PNC} = (-1)^{F_f-M_f}
\left(  \begin{array}{ccc}
                        F_f & 1 & F_i \\
                         -M_f  & q  & M_i \\
                        \end{array} \right )
                                               \langle J_fF_f \| d_{PNC} \| J_iF_i\rangle.
\end{align}
By means of the standard angular momentum technique, the matrix element of $h_Q$ between the hyperfine states $|(JI) F M\rangle$ and $|(J'I) F' M'\rangle$  can be written as a product of the reduced matrix elements of the electronic part and the nuclear part of the interaction:
\begin{align}
\label{eq:quad_matrix}
\langle (J' I) F' M'|h_Q|(JI) F M\rangle \propto \delta_{F'F} \delta_{M'M}(-1)^{F+J+I}\nonumber\\
\left\{  \begin{array}{ccc}
                        J' & J & 2 \\
                         I  & I  & F \\
                        \end{array} \right \}
                                               \langle J' \|  \mbox{{\boldmath $T$}}\| J\rangle\langle I \| \mbox{{\boldmath $Q$}}^{TW} \| I \rangle \,.
\end{align}
The formula for the reduced matrix element of the PNC amplitude induced  by the weak quadrupole $Q^{TW}$ can be derived similar to the derivation of the nuclear-spin-dependent (SD) PNC amplitude in Refs.\cite{PK01} and \citep{JSS03}. The result is
\begin{widetext}
\begin{align}
\label{eq:Q_PNC} 
\langle J_fF_f \| d_Q \| J_iF_i\rangle=\sqrt{\frac{(2I+3)(2I+1)(I+1)}{I(2I-1)}}\sqrt{[F_i][F_f]}
\sum_n \Bigg[(-1)^{J_f-J_i}\left\{\begin{array}{ccc}
                                                J_n & J_f & 1\\
                                                 F_f & F_i & I\\
                                                 \end{array}\right \}\left\{\begin{array}{ccc}
                                                 J_n & J_i & 2\\
                                                  I & I & F_i \\
                                                  \end{array}\right \}\nonumber\\
                                        \times\frac{\langle J_f\|\mbox{{\boldmath $d$}}\|nJ_n\rangle\langle nJ_n\|\mbox{{\boldmath $h$}}_Q^e\|Ji\rangle}{E_n-E_i}
                                                  +(-1)^{F_f-F_i}\left\{\begin{array}{ccc}
                                                J_n & J_i & 1\\
                                                 F_i & F_f & I\\
                                                 \end{array}\right \}\left\{\begin{array}{ccc}
                                                 J_n & J_f & 2\\
                                                  I & I & F_f \\
                                                  \end{array}\right \} \frac{\langle J_f\|\mbox{{\boldmath $h$}}_Q^e\| n J_n\rangle\langle nJ_n\|\mbox{{\boldmath $d$}}\|Ji\rangle}{E_n-E_f}\Bigg].
\end{align}
\end{widetext}
Here $$\mbox{{\boldmath $h$}}_Q^e= - \frac{5 G_F Q^{TW}}{4\sqrt{2}  \langle r^2 \rangle }\mbox{{\boldmath $C$}}^{(2)}\gamma_5 \rho_0(r)$$ is the electronic tensor part of the weak  interaction and  the notation  $[F_a]\equiv 2F_a+1$ is used.  

For comparison we present  two other contributions to the PNC amplitudes in atoms, namely the nuclear spin independent (SI) weak charge $Q_W$ contribution and the nuclear spin dependent (SD) contribution dominated by the magnetic interaction of atomic electrons with the nuclear  anapole moment (AM)\citep{FK80,FKS84}. They have been measured and calculated in many atomic systems - see e.g.  \cite{WBCMRTW97,BJS90,KPT01,PBD09,DBFR12,MZWSH91,WTS93,
MVMLF93,PEBN96,VMMLF95,EPBN95,TDFSYB09,
GCJJLSWBPS03,GLB05,DFSS88,DFSS87,KPJ01,DFa11,NSFK77,DFG02,NK75,DF11}.   
The reduced matrix elements of SI and SD PNC amplitudes are presented e.g. in Ref.  \cite{DFH11}:
\begin{widetext}
\begin{align}
\label{eq:SD_PNC} 
\langle J_fF_f \| d_{SD} \| J_iF_i\rangle=\sqrt{\frac{(2I+1)(I+1)}{I}}\sqrt{[F_i][F_f]}
\sum_n \Bigg[(-1)^{J_f-J_i}\left\{\begin{array}{ccc}
                                                J_n & J_f & 1\\
                                                 F_f & F_i & I\\
                                                 \end{array}\right \}\left\{\begin{array}{ccc}
                                                 J_n & J_i & 1\\
                                                  I & I & F_i \\
                                                  \end{array}\right \}\nonumber\\
    \times           \frac{\langle J_f\|\mbox{{\boldmath $d$}}\|nJ_n\rangle\langle nJ_n\|\mbox{{\boldmath $h$}}_{SD}\|Ji\rangle}{E_n-E_i}
                                                  +(-1)^{F_f-F_i}\left\{\begin{array}{ccc}
                                                J_n & J_i & 1\\
                                                 F_i & F_f & I\\
                                                 \end{array}\right \}\left\{\begin{array}{ccc}
                                                 J_n & J_f & 1\\
                                                  I & I & F_f \\
                                                  \end{array}\right \} \frac{\langle J_f\|\mbox{{\boldmath $h$}}_{SD}\| n J_n\rangle\langle nJ_n\|\mbox{{\boldmath $d$}}\|Ji\rangle}{E_n-E_f}\Bigg]
\end{align}
\end{widetext}
where the vector operator  is the electronic part of the   SD interaction $\mbox{{\boldmath $h$}}_{SD}=(G_F/\sqrt{2})\kappa\mbox{{\boldmath $\alpha$}}\rho_0(r)$ and the
Dirac matrix is defined by
$\mbox{{\boldmath $\alpha$}}=\left(\begin{array}{cc} 0 & \mbox{{\boldmath $\sigma$}}\\
	                                    \mbox{{\boldmath $\sigma$}} & 0 
	        \end{array} \right)\nonumber$. The dimensionless parameter $\kappa$ determines the  strength of the SD PNC interaction . The three major contributions to  $\kappa$ come from the electromagnetic interaction of the atomic electrons with the nuclear anapole moment \cite{FK80,FKS84}, the electron-nucleus SD weak interaction  \cite{NSFK77}, and the combined effect of the SI weak interaction and the magnetic hyperfine interaction \cite{FK85}.
	        
For the SI PNC reduced amplitude we have 
\begin{widetext}
\begin{align}
\label{eq:SI_PNC} 
\langle J_fF_f \| d_{SI} \| J_iF_i\rangle=(-1)^{I+F_i+J_f}\sqrt{[F_i][F_f]}\left\{\begin{array}{ccc}
                                                J_i & J_f & 1\\
                                                 F_f & F_i & I\\
                                                 \end{array}\right \}
\sum_n \Bigg[ \frac{\langle J_f\|\mbox{{\boldmath $d$}}\|nJ_n\rangle\langle nJ_n|H_{SI}|Ji\rangle}{E_n-E_i}
                                                  + \frac{\langle J_f|H_{SI}| n J_n\rangle\langle nJ_n\|\mbox{{\boldmath $d$}}\|Ji\rangle}{E_n-E_f}\Bigg].
\end{align}
\end{widetext}
where the weak interaction is $ H_{SI}=-G_FQ_W/(2\sqrt{2})\gamma_5\rho_0(r)$ and  $Q_W$ is the nuclear weak charge. Note that the weak matrix elements $\langle nJ_n|H_{SI}|Ji\rangle$ are not reduced ones in Eq.(\ref{eq:SI_PNC}).

The single-electron orbitals used to calculate the matrix elements  are
\begin{eqnarray}
\label{diracspi}
\varphi_{n\kappa m}({\bf r}) = \frac{1}{r}\left(   { f_{n\kappa}(r)\Omega_{\kappa m}(\theta,\phi) \atop i g_{n\kappa}(r)\Omega_{-\kappa m}(\theta,\phi)}\right),
\end{eqnarray}
where $n$ is the principle quantum number and  $\kappa=\mp (j+1/2)$ (for $j=l\pm 1/2$) is the angular quantum number for the Dirac spinor.  The relativistic single-particle matrix elements of the  PNC operators are:
 \begin{align}
\label{eq:Q_single} 
\langle \kappa_1\| \mbox{{\boldmath $h$}}^e_Q \|\kappa_2\rangle=i\frac{5G_F Q^{TW}}{4\sqrt{2} \langle r^2 \rangle}\langle \kappa_1\| C^{(2)}\|-\kappa_2\rangle\nonumber\\
\times\int(f_1g_2-g_1f_2)\rho_0~ dr
\end{align}
 \begin{align}
\label{eq:SD_single} 
\langle \kappa_1\| \mbox{{\boldmath $h$}}_{SD} \|\kappa_2\rangle=-i \frac{G_F\kappa}{\sqrt{2}}\langle \kappa_1\| C^{(1)}\|\kappa_2\rangle\nonumber\\
\times\int[(\kappa_1-\kappa_2+1)g_1f_2-(\kappa_2-\kappa_1+1)f_1g_2)\rho_0~ dr
\end{align}
\begin{align}
\label{eq:SI_single} 
\langle \kappa_1|  H_{SI} |\kappa_2\rangle=i \frac{G_FQ_W}{2\sqrt{2}}\delta_{ -\kappa_1,\kappa_2}
\int(f_1g_2-g_1f_2)\rho_0~ dr
\end{align}
Note that all weak matrix elements have  imaginary values.
The reduced matrix element of $\mbox{{\boldmath $C$}}^{(k)}$  is given by
 \begin{align}
\label{eq:Ck_single} 
\langle \kappa_1\| \mbox{{\boldmath $C$}}^{(k)} \|\kappa_2\rangle = (-1)^{j_2+\frac{1}{2}}\sqrt{[j_1][j_2]}\nonumber\\
\xi(l_1+l_2+k)\left (\begin{array}{ccc}
                                                j_2 & j_1 & k\\
                                                -1/2 & 1/2 & 0\\
                                                 \end{array}\right )
\end{align}
where $\xi(L)=1$ if $L$ is even number, otherwise it is zero.


\begin{thebibliography}{99}

\bibitem{K91} I. B. Khriplovich,{\it Parity Nonconservation in Atomic Phenomena}(Gordon and Breach, New York, 1991).

\bibitem{GF04} J. S. M. Ginges and V. V. Flambaum, Phys. Rep.{\bf 397}, 63 (2004).

\bibitem{WBCMRTW97} C. S. Wood, S. C. Bennett, D. Cho, B. P. Masterson, J. L. Roberts, C. E. Tanner, and C. E. Wieman, Science {\bf 275}, 1759 (1997).

\bibitem{D89} V.A. Dzuba, V.V. 	Flambaum, O.P. Sushkov, Phys. Lett. A 141,147, 1989.

\bibitem{BJS90} S. A. Blundell, W. R. Johnson, and J. Sapirstein, Phys. Rev. Lett. {\bf 65}, 1411 (1990).

\bibitem{KPT01} M. G. Kozlov, S. G. Porsev, and I. I. Tupitsyn, Phys. Rev. Lett. {\bf 86}, 3260 (2001).

\bibitem{DFG02} V. A. Dzuba, V. V. Flambaum, and J. S. M. Ginges, Phys. Rev. D {\bf 66}, 076013 (2002).

\bibitem{PBD09} S. G. Porsev, K. Beloy, and A. Derevianko, Phys. Rev. Lett {\bf 102}, 181601 (2009).

\bibitem{DBFR12} V. A. Dzuba, J. C. Berengut, V. V. Flambaum, and B. M. Roberts, Phys. Rev. Lett. {\bf 109}, 203003 (2012).

\bibitem{FK80}  V.V. Flambaum, I.B. Khriplovich. Zh.Exp.Teor.Fiz.  {\bf 79}, 1656, 1980.  [Sov. Phys. JETP {\bf 52}, 835 (1980)].

\bibitem{FKS84}   V.V. Flambaum, I.B. Khriplovich, O.P. Sushkov. Phys. Lett. B. {\bf 146}, 367 (1984).

\bibitem{VMMLF95} P. A. Vetter, D. M. Meekhof, P. K. Majumder, S. K. Lamoreaux, and E. N. Fortson, Phys. Rev. Lett. {\bf 74}, 2658 (1995).

\bibitem{FM97} V. V. Flambaum and D. W. Murray, Phys. Rev. C {\bf 56}, 1641 (1997).

\bibitem{DFK86} V.A.Dzuba, V.V. Flambaum, I.B. Khriplovich, Z. Phys. D {\bf 1}, 243 (1986).

\bibitem{TDFSYB09} K. Tsigutkin,D. R. Dounas-Frazer, A. Family, J. E. Stalnaker, V. V. Yashchuk, and D. Budker, 
Phys. Rev. Lett. {\bf 103}, 071601 (2009).

\bibitem{BDF09} B. A. Brown, A. Derevianko, V. V. Flambaum, Phys. Rev. C {\bf 79}, 035501 (2009).

\bibitem{FS78} O.P. Sushkov, V.V.Flambaum.  Zh.Exp.Teor.Fiz.  {\bf 75}, 1208 (1978)[Sov.Phys. JETP {\bf 48}, 608 (1978)].

\bibitem{KP91} I. B. Khriplovich and M.E. Pospelov, Z. Phys. D {\bf 22}, 367 (1991).

\bibitem{F16} V.V. Flambaum. Phys. Rev. Lett. {\bf 117}, 072501 (2016).

\bibitem{FDK14}V. V. Flambaum, D. DeMille, and M. G. Kozlov, Phys. Rev. Lett. {\bf 113}, 103003 (2014).

\bibitem{DF11} V. A. Dzuba, V. V. Flambaum, Phys. Rev. A {\bf 83}, 052513 (2011).

\bibitem{FrPNC1}
M. Tandecki, J. Zhang, R. Collister, {\em et al.}
J. Instrumentation. {\bf 8}, P12006 (2013).

\bibitem{FrPNC2}
E. Mariotti, A. Khanbekyan, C. Marinelli, {\em et al.}
Int. J. Mod. Phys. E {\bf 23}, 1430009 (2014).

\bibitem{YbPNC} V. A. Dzuba and V. V. Flambaum,                            
      Phys. Rev. A {\bf 83}, 052513 (2011).

\bibitem{RaPNC}
M. Nunez Portela, E.A. Dijck, A. Mohanty,  {\em et al.}
App. Phys. B {\bf 114}, 173 (2014).

\bibitem{Qp} N. J. Stone, 
Atomic Data and Nuclear Data Tables {\bf 111-112}, 1 (2016).

\bibitem{BiPNC} L. M. Barkov and M. S. Zolotorev, Phys. Lett. B {\bf 85}, 308 (1979).

\bibitem{Bikappa} 
V. V. Flambaum, I. B. Khriplovich, and O. P. Sushkov, Phys. Lett. B {\bf 146}, 367 (1984).

\bibitem{NIST} Kramida, A., Ralchenko, Yu., Reader, J., and NIST ASD
  Team (2013). NIST Atomic Spectra Database (ver. 5.1),
  [Online]. Available: http://physics.nist.gov/asd [2013, December
  12]. National Institute of Standards and Technology, Gaithersburg,
  MD. 

\bibitem{DyPNC}
A.T. Nguyen, D. Budker, D. DeMille,  and M. Zolotorev,  Phys. Rev. A {\bf 56}, 3453 (1997).

\bibitem{Ra-Ginges}  V.A.Dzuba, V.V.Flambaum and J.S.M. Ginges,
      Phys. Rev. A, {\bf 61}, 062509 (2000).

\bibitem{DFH11} V. A. Dzuba, V. V. Flambaum, and C. Harabati, Phys. Rev. A {\bf 84}, 052108 (2011).

\bibitem{PK01} S. G. Porsev and M. G. Kozlov, Phys. Rev. A {\bf 64}, 064101 (2001).

\bibitem{JSS03} W. R. Johnson, M. S. Safronova, U. I. Safronova, Phys. Rev. A {\bf 67}, 062106 (2003).





\bibitem{FK85} V. V. Flambaum and I. B. Khriplovich, ZhETP {\bf 89}, 1505 (1985)[Soviet Phys. JETP {\bf 62}, 872 (1985)].

\bibitem{MZWSH91} M. J. D. Macpherson, K. P. Zetie, R. B. Warrington, D. N. Stacey, and J. P. Hoare, Phys. Rev. Lett. {\bf 67}, 2784 (1991).

\bibitem{WTS93} R. B. Warrington, C. D. Thompson, and D. N. Stacey, Europhys. Lett. {\bf 24}, 641 (1993).

\bibitem{MVMLF93} D. M. Meekhof, P. A. Vetter, P. K. Majumder, S. K. Lamoreaux, and E. N. Fortson, Phys. Rev. Lett. {\bf 71}, 3442 (1993).

\bibitem{PEBN96} S. J. Phipp, N. H. Edwards, P. E. G. Baird, and S. Nakayama, J. Phys. B {\bf 29}, 1861 (1996).


\bibitem{EPBN95} N. H. Edwards, S. J. Phipp, P. E. G. Baird, and S. Nakayama, Phys. Rev. Lett. {\bf 74}, 2654 (1995).


\bibitem{GCJJLSWBPS03} J. Gu\'{e}na, D. Chauvat, P. Jacquier, E. Jahier, M. Lintz, S. Sanguinetti, A. Wasan, M.-A. Bouchiat, A. V. Papoyan, and D. Sarkisyan, Phys. Rev. Lett. {\bf 90}, 143001 (2003).

\bibitem{GLB05} J. Gu\'{e}na, M. Lintz, and M.-A. Bouchiat, Phys. Rev. A {\bf 71}, 042108 (2005).

\bibitem{DFSS88} V. A. Dzuba, V. V. Flambaum, P. G. Silvestrov, and O. P. Sushkov, Europhys. Lett. {\bf 7}, 413 (1988).

\bibitem{DFSS87} V. A. Dzuba, V. V. Flambaum, P. G. Silvestrov, and O. P. Sushkov, J. Phys. B {\bf 20}, 3297 (1987).

\bibitem{KPJ01} M. G. Kozlov, S. G. Porsev, and W. R. Johnson, Phys. Rev. A {\bf 64}, 052107 (2001).

\bibitem{DFa11} V. A. Dzuba, V. V. Flambaum, Phys. Rev. A {\bf 83}, 042514 (2011).

\bibitem{NSFK77} V. N. Novikov, O. P. Sushkov, V. V. Flambaum, and I. B. Khriplovich, ZhETP {\bf 73}, 802 (1977)[Soviet Phys. JETP {\bf 46}, 420 (1977)].

\bibitem{NK75} V. N. Novikov and I. B. Khriplovich, Pis'ma Zh. Eksp. Teor. Fiz. {\bf 22}, 162 (1975)[JETP Lett. {\bf 22}, 74 (1975)].





\end{thebibliography}
\end{document}